# Three-dimensional anisotropic fluctuation diamagnetism around the superconducting transition of $Ba_{1-x}K_xFe_2As_2$ single crystals in the finite-field (or Prange) regime

J. Mosqueira[1,*], J.D. Dancausa[1], C. Carballeira[1], S. Salem-Sugui Jr.[2], A.D. Alvarenga[3], H.-Q. Luo[4], Z.-S. Wang[4], H.-H. Wen[4,5], F. Vidal[1]

[1]*LBTS, Universidade de Santiago de Compostela, ES-15782 Santiago de compostela, Spain*

[2]*Instituto de Fisica, Universidade Federal do Rio de Janeiro, 21941-972 Rio de Janeiro, RJ, Brazil*

[3]*Instituto Nacional de Metrologia Qualidade e Tecnologia,25250-020 Duque de Caxias, RJ, Brazil*

[4]*Beijing National Laboratory for Condensed Matter Physics, Institute of Physics, Chinese Academy of Sciences, Beijing 100190, China*

[5]*Center for Superconducting Physics and Materials, National Laboratory of Solid State Microstructures and Department of Physics, Nanjing University, Nanjing 210093 China*

*j.mosqueira@usc.es, Tel.: +34-981-563100, Fax: +34-881-814 112

**Abstract:**

The magnetization around the superconducting transition was recently measured in a high-quality $Ba_{1-x}K_xFe_2As_2$ single crystal with magnetic fields applied along and transverse to the crystal Fe-layers [J. Mosqueira et al., Phys. Rev. B **83**, 094519 (2011)]. Here we extend this study to the finite field (or Prange) regime, in which the magnetic susceptibility is expected to be strongly dependent on the applied magnetic field. These measurements are analyzed in the framework of the three-dimensional anisotropic Ginzburg Landau (3D-aGL) approach generalized to the short wavelength regime through the introduction of a total-energy cutoff in the fluctuation spectrum. The results further confirm the adequacy of GL approaches to describe the fluctuation effects close to the superconducting transition of these materials.

## 1. Introduction

The discovery of superconductivity in iron pnictides [1] has stimulated an intense research activity in the last few years, not only due to their relatively high transition temperatures, $T_c$, but also to their similarities with the high-$T_c$ cuprates (a layered crystal structure, a similar evolution of the superconducting parameters with doping, the proximity to a magnetic transition, and a pairing mechanism possibly mediated by magnetic interactions).[2] An aspect of the physics of these materials which still received very little attention is the diamagnetism induced by superconducting fluctuations above $T_c$. This effect, also called *precursor diamagnetism*, is a very useful tool to probe the nature of a superconducting transition and to obtain material parameters as the upper critical fields or the anisotropy factor.[3] In the case of iron-pnictides, the only work on the subject was the above mentioned one [4], performed in a $Ba_{1-x}K_xFe_2As_2$ single crystal (x=0.28) with magnetic fields applied both parallel and perpendicular to the *ab* (Fe) layers. It was shown that the three-dimensional anisotropic Ginzburg-Landau approach (3D-aGL) with a total-energy cutoff (Ref. 5) explained at a quantitative level the experimental data in the low-field limit. Here we extend this study to the finite-field (or Prange) regime, in which the fluctuation magnetic susceptibility is expected to present a markedly field-dependent behavior.[3,6]

## 2. Experimental details and results

The $Ba_{1-x}K_xFe_2As_2$ single crystal used in this work is the same as in Ref. 4. Details of its growth and characterization may be found in Ref. 7. Let us just mention that it has a well defined low-field diamagnetic transition with $T_c$ = 33.2 ± 0.2 K (see Ref. 4), which allows to determine fluctuation effects down to very close to the transition temperature, for $[T-T_c(H)]/T_c(H) \approx 10^{-2}$. The measurements were performed with a SQUID magnetometer (Quantum Design, MPMS-XL), which allows a resolution of $10^{-8}$ emu in magnetic moment, *m*, enough to study fluctuation effects near $T_c$ (in the $10^{-6}-10^{-7}$ emu range in our 0.43 mg crystal). Examples of the as-measured $m(T)$ curves for different magnetic field amplitudes and orientations ($H \perp ab$ and $H // ab$) are shown in Fig. 1. The fluctuation contribution to the magnetic moment was determined from these curves by subtracting the background magnetic moment, $m_B(T)$, coming from the crystal normal state and in a much lesser extent from the quartz sample holder used. Such a background contribution was determined by fitting a Curie-like function, $m_B(T) = a + bT + c/T$, to the raw data in the temperature interval between 37 and 50 K ($a$, $b$, and $c$ are free parameters). The lower bound corresponds to a temperature above which the fluctuation induced magnetic moment is estimated to be smaller than the experimental uncertainty [4]. The resulting fluctuation magnetic susceptibility $\Delta M/H=(m-m_B)/VH$ is presented in Fig. 2. As it may be clearly seen, when $H \perp ab$ the rounding associated to fluctuations is more pronounced and the sharp $\Delta M/H$ drop observed on approaching $T_c(H)$ is more strongly shifted to lower temperatures, both effects being due to the slight anisotropy of the compound studied.[4] An example (corresponding to $T = T_c$) of the magnetic field dependence of $\Delta M/H$ for both field orientations is presented in the inset in Fig. 2(a). As it may be clearly seen, $\Delta M/H$ presents a strong dependence with $H$, in contrast with the $H$-independent behavior observed in the low-field limit, when $h << \varepsilon$ [4].

## 3. Data analysis

To analyze quantitatively the results of Fig. 2, as the $c$-axis superconducting coherence length is larger than the layers periodicity length,[4] we will use the 3D anisotropic Gaussian Ginzburg-Landau (3D-aGGL) theory. Buzdin and Feinberg derived an expression for $\Delta M$ in the framework of this approach for magnetic fields with an arbitrary amplitude and orientation.[9] However, their result is not applicable in the short-wavelength fluctuation regime, i.e. at high reduced temperatures $\varepsilon \equiv \ln(T/T_c)$ or high reduced magnetic fields $h \equiv H/H_{c2}^\perp(0)$, where $H_{c2}^\perp(0)$ is the upper critical field for $H \perp ab$ extrapolated to 0 K. In Ref. 5 it was shown that the introduction of a *total-energy cutoff* in the fluctuation spectrum extends the applicability of GGL approaches to these short wavelength regimes. In particular, it explains the observed vanishing of fluctuation effects at $\varepsilon \sim 0.5$ in the low-field limit [5], and h ~ 1 when T → 0 K [9]. By introducing the total-energy cutoff in the 3D-aGL approach for the fluctuation magnetization, for $H \perp ab$ it was obtained [10,11]

$$\Delta M_\perp(\varepsilon, h) = -\frac{k_B T \gamma \sqrt{2h}}{\pi \phi_0 \xi_{ab}(0)} \int_0^{\sqrt{(c-\varepsilon)/2h}} dx \left[ \frac{c-\varepsilon}{2h} + \left(\frac{\varepsilon}{2h} + x^2\right) \psi\left(\frac{\varepsilon+h}{2h} + x^2\right) - \ln \Gamma\left(\frac{\varepsilon+h}{2h} + x^2\right) - \right.$$

$$\left. \left(\frac{c}{2h} + x^2\right) \psi\left(\frac{c+h}{2h} + x^2\right) + \ln \Gamma\left(\frac{c+h}{2h} + x^2\right) \right]. \tag{1}$$

Here $\Gamma$ and $\psi$ are, respectively, the gamma and digamma functions, $\gamma$ the anisotropy factor, $\xi_{ab}(0)$ the in-plane coherence length amplitude, and $c = 0.55$ the cutoff constant [5]. In the limit $c \to \infty$ (i.e., in the absence of cutoff) Eq. (1) leads to

$$\Delta M_\perp = -\frac{k_B T \gamma \sqrt{2h}}{\pi \phi_0 \xi_{ab}(0)} \int_0^\infty dx \left\{ \left(\frac{\varepsilon}{2h} + x^2\right) \left[ \psi\left(\frac{\varepsilon+h}{2h} + x^2\right) - 1 \right] - \ln \Gamma\left(\frac{\varepsilon+h}{2h} + x^2\right) + \ln \sqrt{2\pi} \right\}, \tag{2}$$

which for isotropic materials ($\gamma = 1$) is equivalent to the one derived by Prange [Eq. (3) in Ref. 6, for details see Ref. 10]. In the low magnetic field limit $h \ll \varepsilon$, Eq. (1) reduces to

$$\Delta M_\perp(\varepsilon, h) = -\frac{k_B T \gamma h}{6\pi \phi_0 \xi_{ab}(0)} \left[ \frac{\tan^{-1}\sqrt{(c-\varepsilon)/\varepsilon}}{\sqrt{\varepsilon}} - \frac{\tan^{-1}\sqrt{(c-\varepsilon)/c}}{\sqrt{c}} \right], \tag{3}$$

which is linear in $h$. In the absence of any cutoff it further simplifies to

$$\Delta M_\perp(\varepsilon, h) = -\frac{k_B T \gamma h}{12 \phi_0 \xi_{ab}(0)} \varepsilon^{-1/2}, \tag{4}$$

which for isotropic materials is the well known Schmidt result [3,12]. The fluctuation magnetization for $H \parallel ab$ for finite magnetic fields may be related to the one for $H \perp ab$ according to [13]

$$\Delta M_\parallel(\varepsilon, h) = \frac{1}{\gamma} \Delta M_\perp\left(\varepsilon, \frac{h}{\gamma}\right). \tag{5}$$

The lines in Fig. 2(a) correspond to Eq. (1) as evaluated with the same superconducting parameters as in Ref. 4, namely $\mu_0 H_{c2}^\perp(0) = 140$ T and $\gamma = 1.83$, and by using $T_c = 33.3$ K. The slight difference in the $T_c$ value with respect to the one obtained in Ref. 4 (33.2 K) falls within the experimental uncertainty, and may be attributed to the use in Ref. 4 of an expression [Eq.(3)] which is strictly valid when H → 0. As may be seen, the agreement with the experimental data is excellent, and even extends somewhat below $T_c$ [but still above $T_c(H)$]. In turn, Eq. (5) [evaluated with Eq. (1) and with the same superconducting parameters] seems to be in good agreement with the data obtained with $H // ab$, in spite that in this field direction the signal-to-noise ratio is significantly smaller than for $H \perp ab$. As expected, Eqs. (1) and (5) also reproduce the strongly non-linear $H$-dependence of $\Delta M/H$ at $T_c$ presented in the inset of Fig. 2 (solid lines). The behavior of fluctuation effects near $T_c$ in iron pnictides is found to be qualitatively similar to the one observed in high-$T_c$ cuprates [14] and in low-$T_c$ metallic elements and alloys [9,15]. This seems to suggest that, apart from the differences in dimensionality and in the values of the superconducting parameters, fluctuation effects near $T_c$ present similar characteristics in all superconducting families and may be described by conventional GL approaches.

The present experimental results and analysis are a further evidence that the precursor diamagnetism above $T_c(H)$ in the moderately anisotropic 122 iron pnictides may be explained in terms of a conventional GL approach for 3D anisotropic superconductors including a total-energy cutoff in the fluctuation spectrum. This provides a constraint for any microscopic theory for the superconductivity in these materials, and may have implications in systems like the high-$T_c$ cuprates, with similar superconducting parameters and possibly a similar mechanism for their superconductivity. It would be interesting to extend the present measurements to larger reduced magnetic fields in order to enter more deeply in the finite field regime and explore the limits of applicability of Eq. (1). In particular, it would be interesting to check whether fluctuation effects vanish at $h \sim 1.1$ as it was observed in low-$T_c$ alloys [9].

**Acknowledgments**

This work was supported by the Spanish MICINN and ERDF (grant no. FIS2010-19807), by the Xunta de Galicia (grants no. 2010/XA043 and 10TMT206012PR), and the european project ENERMAT. SSS and ADA acknowledge support from the CNPq and FAPERJ.

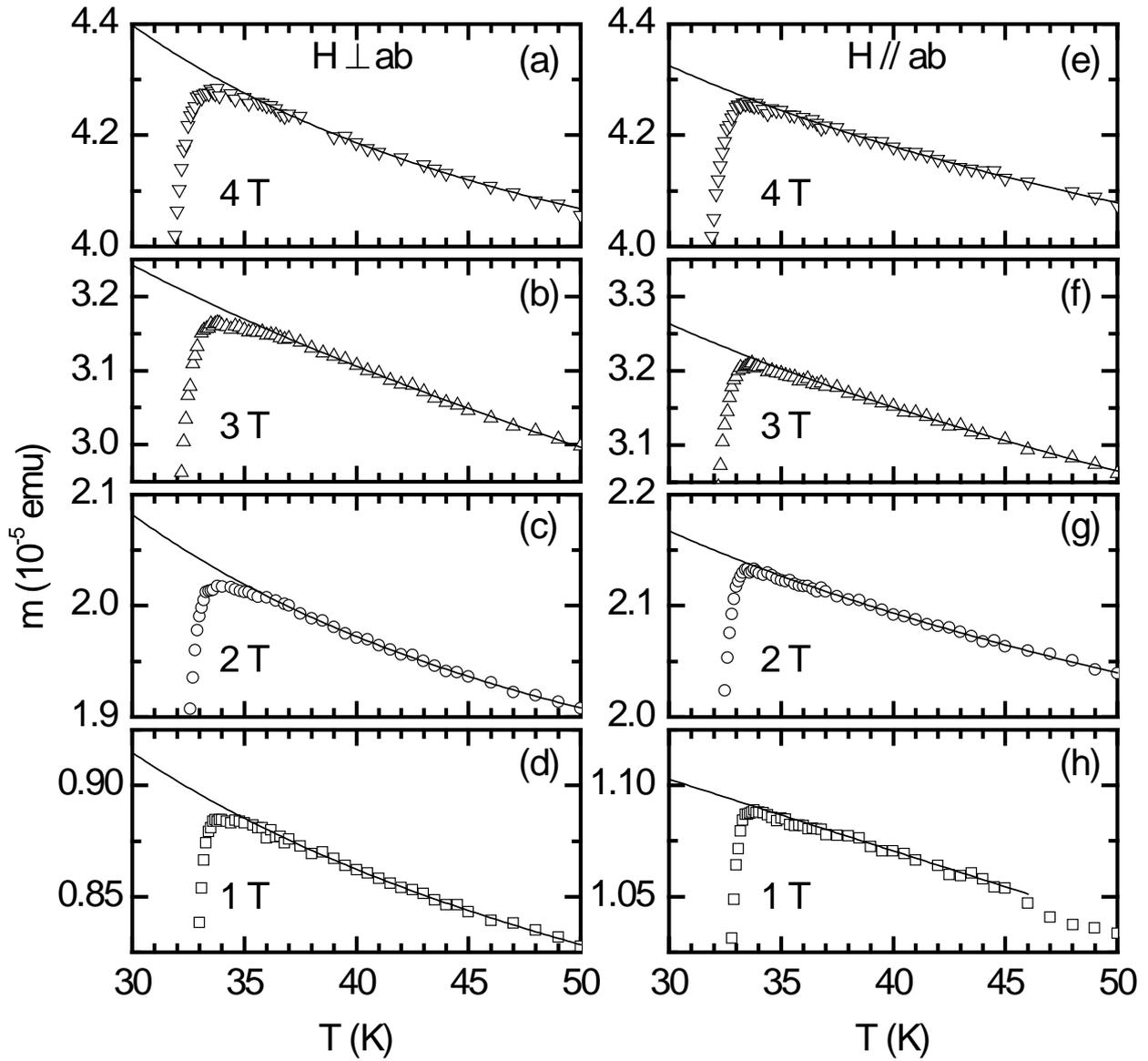

**Fig. 1.** Temperature dependence of the as measured magnetic moment in the normal state for magnetic fields between 1 and 4 T applied perpendicular (a-d) and, respectively, parallel (e-h) to the Fe-layers. The lines are fits of a Curie-like function to the temperature interval between 37 and 50 K. The lower bound corresponds to the temperature at which fluctuation effects become smaller than the instrumental resolution. Such effects are clearly observed as a rounding of the $m(T)$ curves for $H \perp ab$ near $T_c$, but are unobservable in this scale when $H // ab$.

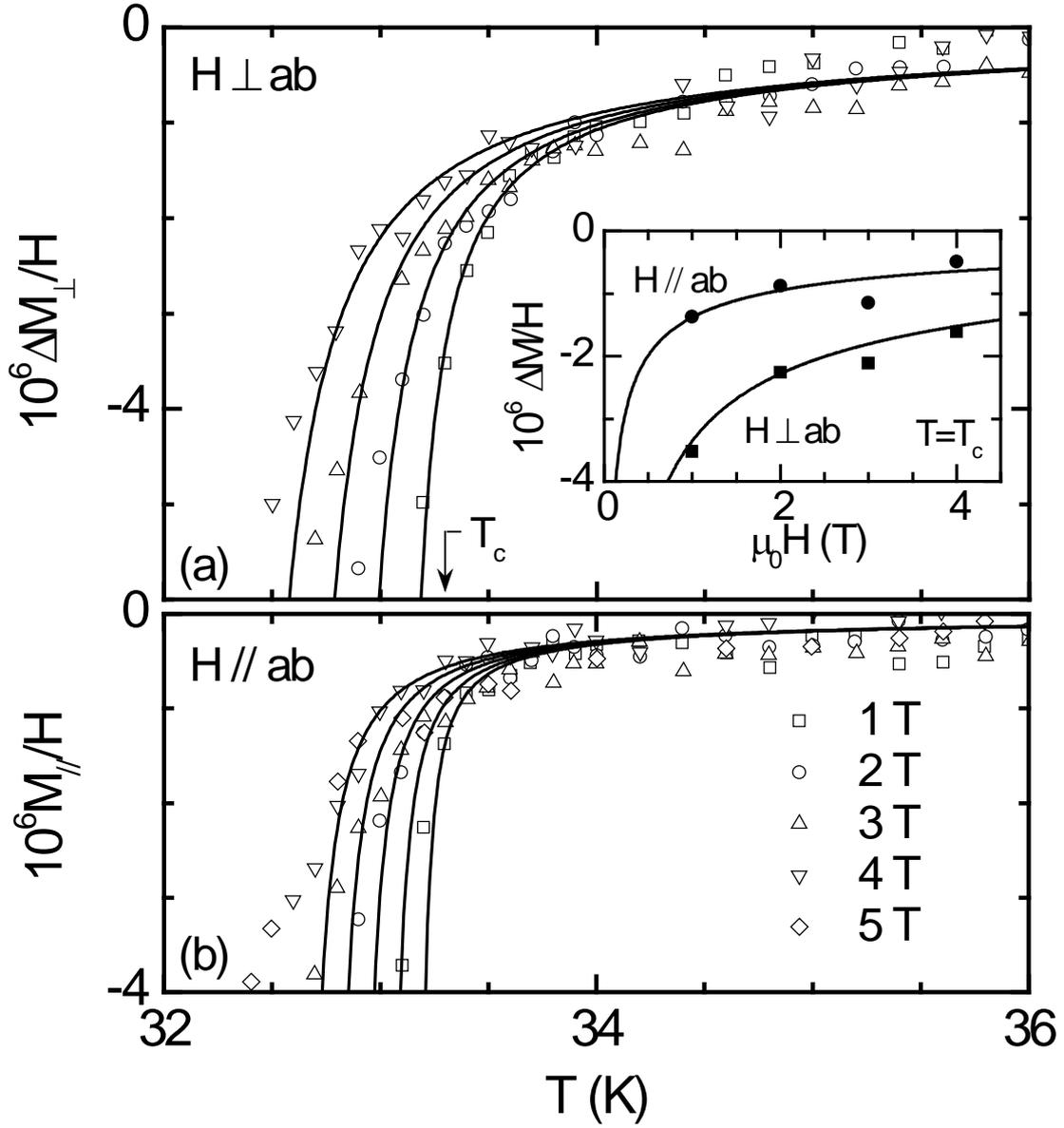

**Fig. 2.** (a) and (b) Temperature dependence of the fluctuation magnetic susceptibility around $T_c$ for $H$ applied in the two main crystal directions. The lines correspond to the 3D anisotropic Ginzburg-Landau result with a total-energy cutoff [Eq. (1) for $H \perp ab$, and Eq. (5) for $H // ab$] evaluated with the $H_{c2}^{\perp}(0)$ and $\gamma$ values derived in Ref. 4. Inset: Isotherms corresponding to $T_c$ with the $H$ dependence of $\Delta M/H$ for both field orientations. The lines correspond to Eqs. (1) and (5) evaluated with the same superconducting parameters.